\documentstyle[12pt]{article}

\newcommand{\be}{\begin{equation}}
\newcommand{\ee}{\end{equation}}
\newcommand{\bea}{\begin{array}}
\newcommand{\ea}{\end{array}}
\newcommand{\beqa}{\begin{eqnarray}}
\newcommand{\eeqa}{\end{eqnarray}}
\newcommand{\bean}{\begin{eqnarray*}}
\newcommand{\eean}{\end{eqnarray*}}

\def\up#1{\leavevmode \raise.16ex\hbox{#1}}

\def\sqr#1#2{{\vcenter{\vbox{\hrule height.#2pt
        \hbox{\vrule width.#2pt height#1pt \kern#1pt
          \vrule width.#2pt}
        \hrule height.#2pt}}}}

\newcommand{\journal}[4]{{\sl #1 }{\bf #2} \up(19#3\up) #4}

\newcommand{\gapproxeq}{\lower .7ex\hbox{$\;\stackrel{\textstyle
>}{\sim}\;$}}
\newcommand{\lapproxeq}{\lower .7ex\hbox{$\;\stackrel{\textstyle
<}{\sim}\;$}}

\newcounter{appendice}

\def\thebibliography#1{{\bf REFERENCES\markboth
 {REFERENCES}{REFERENCES}}\list
 {[\arabic{enumi}]}{\settowidth\labelwidth{[#1]}\leftmargin\labelwidth
 \advance\leftmargin\labelsep
 \usecounter{enumi}}
 \def\newblock{\hskip .11em plus .33em minus -.07em}
 \sloppy
 \sfcode`\.=1000\relax}

\begin{document}
\title{\hfill $\mbox{\small{
$\stackrel{\rm\textstyle DSF-37/99\quad} {\rm\textstyle
hep-th/9912091\quad\quad} $}}$ \\[1truecm]
   Bicovariant Calculus in Quantum Theory and a Generalization of
   the Gauss Law   }
\author{G. Bimonte$^{a}$, G. Marmo$^{a}$ and A. Stern$^{b}$}
\maketitle
\thispagestyle{empty}

\begin{center}
{\it a)  Dipartimento di Scienze Fisiche, Universit\`a di Napoli,\\
Mostra d'Oltremare, Pad.20, I-80125, Napoli, Italy; \\
INFN, Sezione di Napoli, Napoli, ITALY.} \\
 {\it b) Department of Physics, University of Alabama,\\
Tuscaloosa, AL 35487, USA.}

\end{center}

\begin{abstract}

We construct a deformation of the quantum algebra $Fun(T^*G)$ associated
with Lie group $G$    to the case where
 $G$ is replaced by  a quantum group $G_q$ which has a
bicovariant calculus.  The deformation   easily  allows for the
inclusion of the current algebra of  left and right invariant one forms.
We use it  to examine a possible  generalization of the  Gauss
law commutation relations for gauge theories based on $G_q$.

 \end{abstract}

\bigskip

{\bf PACS}:


\newpage

\section{Introduction}

For many descriptions of dynamical systems where
 continuous symmetries are present,
 one requires more than the existence of a Lie
group $G$ and its corresponding Lie algebra $g$.  Rather, one often relies
on the existence of  a bicovariant calculus on $G$.
This is especially true  when constructing currents and current algebras.
A bicovariant calculus is also useful for gauge theories.   Loosely
speaking, the bicovariant calculus allows for
 compatible left and right actions on a bimodule constructed on
 the group.   The bimodule is spanned by
 the left or right invariant  one forms.
There is a natural construction of a bicovariant calculus for any
Lie group  $G$.

The same cannot be said for  quantum groups.
 The notion of  a bicovariant calculus
 for  quantum groups  was introduced by    Woronowicz \cite{bico}.
 (Also see \cite{db}, \cite{ju},
 \cite{wsww}, \cite{swa}, \cite{ac}, \cite{ss}).)
Only for a class of  quantum groups $\{ G_q\}$
 does a structure analogous to the one
 for Lie groups exist.
      There have  subsequently been several
   attempts  to apply this structure to  dynamical systems
\cite{Cas},\cite{us}.  In these works, one finds
 generalizations of
 the usual classical Lagrangians for gauge theories, including
  Chern-Simons theory and gravity.  However, such systems have the novel
feature that  the `classical'  field degrees of freedom end up
 being noncommuting operators.
Their physical interpretation is, to say the least, obscure.
The situation is somewhat reminiscent of the `classical' Lagrangian
description for fermions expressed in terms of Grassmann fields.
That description derives a meaning due to the existence of a canonical
quantization scheme which turns it into the usual quantum mechanical
 description of fermions.  An analogous canonical quantization scheme
 for `classical' Lagrangian descriptions of
 quantum group gauge symmetries is not obvious.
 This being the case, there is currently
not much use  in having such  Lagrangians, aside
from being a curiosity.

Under these circumstances it is
reasonable to skip the step of canonical quantization entirely,
and start instead  with the quantum theory.
  That is, we can try to incorporate
the structure of a bicovariant calculus for a quantum group $G_q$ in
quantum theory.       For ordinary Lie groups $G$,
 a natural setting for a bicovariant calculus
   is provided by  the cotangent bundle $T^*G$.   Left and right actions
on $T^*G$ are obtained  via the pullback from $G$ of right invariant or
left invariant one-forms, and
 they are generated using the corresponding
canonical momenta coordinatizing the fibre, via the trivialization
 $T^*G=G
\otimes g^*$.  ($g^*$ is the co-Lie algebra associated with $G$.)  In
the quantum theory, one defines
 the algebra of functions $Fun(T^*G)$  on  $T^*G$.    This is the
algebra of operators for the rigid rotor, and it is generated by
 elements of  $G$ and their conjugate momenta.  The algebra is easily
enlarged to include the current algebra of the
 left and right invariant one forms.

Our task is  to find i) a deformation of $Fun(T^*G)$ to an algebra
${\cal  B}$ where $G$ is replaced by   $G_q$, and then ii)
enlarge that algebra to include the left and right invariant one forms.
For the latter, the existence of an exterior derivative is assumed
with the usual Leibniz rule.  (For an alternative to
 the usual Leibniz rule, see
\cite{fp}.)
We shall carry out i) and ii) for quantum groups $G_q$ having
 a bicovariant calculus.  With regard to i),     deformations of
$Fun(T^*G)$ known as   Heisenberg doubles exist
 for a large class of quantum groups,
 which do not necessarily possess a  bicovariant calculus.\cite{sts}
  We instead find it convenient to write down a deformation algebra
    ${\cal B}$
exclusively for quantum groups which posses a bicovariant calculus,
as it is easily  extended  to the bimodule.  A bicovariant calculus is
  characterized by braiding matrices, structure constants, as well as
the $R$ matrices for $G_q$, and we express
${\cal B}$ directly in terms of these constants.
   The algebra is shown to be consistent with
 associativity, and although  it is not in general a Hopf algebra,
 it contains three Hopf subalgebras:  the  quantum group
 $G_q$, as well as two algebras ${\cal A}_L$ and ${\cal A}_R$ which are
deformations of the Lie algebra associated with $G$.
  ${\cal A}_L$ and ${\cal A}_R$ contain the analogues of
  generators of the left and right actions of Lie groups, which we
denote by    $\{\ell_i\}$ and $\{r_i\}$, respectively.  These algebras
commute with each other, and they may or may not be isomorphic.
  (We find them to be isomorphic for simple examples, like $G_q=U_q(2)$.)
   Their center elements are shown  to coincide.
All of the above properties of ${\cal B}$ are shared by
the Heisenberg double\cite{sts}
  \footnote{We thank S. Frolov for bringing this to our attention.},
and most likely any ${\cal B}$ is equivalent to some quantum double.
As stated earlier,  using  ${\cal B}$  we can easily
carry out step ii), i.e.
extend the algebra to include  the left and right invariant one forms.
In this regard we get that  $\{\ell_i\}$
   commutes with the left  invariant one form, and $\{r_i\}$
   commutes with the right invariant one form, as in the undeformed case.
The remaining commutation relations can be expressed in terms of
deformed commutators.

The work presented
here sets the stage for a variety of generalizations
of quantum systems, such as  Wess-Zumino-Witten models
and gauge theories, to ones based on quantum groups.   With regard to
the latter, we propose a  simple generalization of the  Gauss
law commutation relations.  We find that the resulting algebra
is associative only for a certain restricted class of deformations,
known as `minimal'\cite{Cas}.   For such
deformations we express the Gauss law in terms of the analogue
of potentials and electric fields and specify their algebra.
 We plan to explore its  consequences  in  future works.

We first review the
 algebra  $Fun(T^*G)$ of
functions on the cotangent bundle $T^*G$ associated with Lie group $G$.
It is generated by  group elements, along with elements
of the Lie algebra.   For the former one can write matrices
 $[T^a_{\;\;b}]$ in the fundamental representation.
 \footnote{We restrict here our attention to subgroups of the general
linear group.}
 The latter, which we denote by  $r_i$ (or $\ell_i$)
are  the generators  of
 right  (or left) transformations of $G$ on $T^*G$ and project onto
corresponding
  actions on the group $G$. (The indices $a,b,..$
label the fundamental representation, while $i,j,..$ are Lie algebra
indices.)   In terms of the right generators, the algebra is  given by
\beqa                                              \label{TTud}
[T^a_{\;\;b}, T^c_{\;\;d}]& =& 0        \\          \label{rrud}
 [r_i,r_j] &=& c_{ji}^k \;  r_k        \\          \label{Trud}
[T^a_{\;\;b}, r_i] & =& T^a_{\;\;c}\; [\lambda_i]^c_{\;\;b}      \eeqa
where  $ c_{ij}^k $ are structure constants and
 $\{ [\lambda_i]^a_{\;\;c}\}$ spans the
   fundamental representation of the Lie algebra, i.e.
$  [\lambda_i,\lambda_j]  =  c_{ij}^k \;  \lambda_k  $.  Eqs.
(\ref{TTud}-\ref{Trud}) define  the quantum algebra for a rigid rotor
(where here we set $\hbar=1$).
    The left generators $\ell_i$  can be expressed in
terms  of the right generators
 upon introducing adjoint matrices $M^{\;i}_j$ satisfying
\be \lambda_j T = T\lambda_i \; M^{\;i}_j \label{defM} \ee
 Their commutation relations are
\beqa \label{MM} [M^{\;i}_j, M^{\;k}_\ell ] &=& 0  \\   \label{Mr}
[M^{\;i}_j, r_k] & =&  c^i_{\ell k}M^{\;\ell}_j  \eeqa
Defining  $\ell_i=r_j M^{\;j}_i$, gives
\beqa \label{lrcom}  [ \ell_i , r_j ] &=& 0     \\
\label{llud}     [\ell_i , \ell_j ] &=& c_{ij}^k \ell_k  \\
\label{Tellud}   [T^a_{\;\;b}, \ell_i] & =&    [\lambda_i]^a_{\;\;c}
 T^c_{\;\;b}     \eeqa    where we used  \be
c_{ij}^k  M^{\;\ell}_k=M^{\;k}_i M^{\;n}_j c_{kn}^\ell \label{cMM}\ee
Upon introducing the left and right invariant one forms:
   \be \omega_L = T^{-1}dT \;,\qquad
       \omega_R = dT T^{-1}  \;,\label{lrif} \ee
 it is easy to check that $\omega_L$ commutes with
$\ell_i$, and $\omega_R$ commutes with $r_i$.
The nonvanishing commutation relations of the generators
 $\ell_i$ and $r_i$ with
 $\omega_L = \omega_L^i \;\lambda_i$  and
             $\omega_R = \omega_R^i \;\lambda_i$ are
\beqa           [\omega_L^i,  r_j] &=& c^i_{kj}\; \omega_L^k
      \cr      [\ell_j,\omega_R^i]&=& c^i_{kj}\; \omega_R^k
         \label{crowlrud}   \eeqa
The components $ \omega_L^i$ and $\omega_R^i$ are related by
\be  \omega_L^i = \omega_R^j  M^{\;i}_j  \label{rbos}   \ee

We now show that there is a completely analogous structure for quantum
 groups $G_q$ which admit a bicovariant calculus in the sense of
 Woronowicz\cite{bico}.
 Quantum groups can be thought of as deformations of the
(commutative) Hopf
algebra $Fun(G)$ of functions on a Lie group $G$. The deformation  of
$Fun(G)$  to a noncommutative Hopf algebra $G_q$ depends on at least one
deformation parameter $q$, and
reduces to $Fun(G)$ in the limit  $q\rightarrow 1$.
The Hopf algebra is generated by
 quantum group matrices for which we again use the notation
  $ T$.   Their coproduct is the usual one:
$\Delta( T^a_{\;\;b}) =  T^a_{\;\;c} \otimes T^c_{\;\;b}$, and
   they satisfy commutation relations:
\be R^{ab}_{\;\;\;ef}\;T^e_{\;\;c} T^f_{\;\;d}- T^b_{\;\;f}
T^a_{\;\;e}\;
 R^{ef}_{\;\;\;cd} = 0 \label{RTT}
\ee
$R$ obeys the Yang-Baxter equation, and this is consistent with
having an associative algebra for the matrix elements $T^a_{\;\;b}$.
  We want  to enlarge this
algebra to an associative algebra ${\cal B}$ which is a  deformation of
  $Fun(T^*G)$.  For this we  reintroduce
elements $r_i$, which now are  the analogues of the  right generators on
the group.  For their commutation
 relations we take:
\beqa
     r_i r_j -\Lambda^{mn}_{ji} r_n r_m &=&c^k_{ji} r_k \label{rr} \\
      T^a_{\;\;b} r_i - r_j T^a_{\;\;c} \;[f^j_i]^c_{\;\;b} &=&
    T^a_{\;\;c}  [\chi_i]^c_{\;\;b} \label{Tr}
      \eeqa      which, along with (\ref{RTT}),
generalize  (\ref{TTud}-\ref{Trud}).   Eq. (\ref{rr}) defines a
deformed Lie algebra.
$\Lambda^{km}_{ij}$ and  $c_{ij}^k$ are c-numbers,
 which are respectively, the braiding matrices
and structure constants.   For $q\rightarrow 1$,  $\Lambda^{km}_{ij}
\rightarrow \delta^k_j\delta^m_i$.  For $q \ne 1$,
the  structure constants are not in general antisymmetric, and do not
 obey the usual
Jacobi identities.  In systems where there is
  a bicovariant calculus, one has the following identities for
$\Lambda^{km}_{ij}$ and  $c_{ij}^k$\cite{ac}:
\beqa
\Lambda^{ij}_{kl}\Lambda^{lm}_{sp} \Lambda^{ks}_{qu} =
\Lambda^{jm}_{kl}\Lambda^{ik}_{qs} \Lambda^{sl}_{up}~~~
 \label{yb} \\
c^r_{mi} c^n_{rj}  -\Lambda^{kl}_{ij} c^r_{mk} c^n_{rl} =
c_{ij}^k c_{mk}^n~~~ \label{qjac} \\
\Lambda^{ir}_{mk}\Lambda^{ks}_{nl} c_{rs}^j = \Lambda^{ij}_{kl} c_{mn}^k
~,       \label{2d} \\
\Lambda^{jq}_{ri}\Lambda^{si}_{kl} c_{ps}^r + \Lambda^{jq}_{pi}
c_{kl}^i=
 c_{is}^j \Lambda^{sq}_{rl}\Lambda^{ir}_{pk} +
c_{rl}^q\Lambda^{jr}_{pk}~
 \label{id4}\eeqa
The first condition is the Yang Baxter equation, the second
is the analogue of the Jacobi identity, while the
last two equations are trivial in the limit $q\rightarrow 1$.
 From the above conditions   the $r_i$'s (alone)
 generate a non cocommutative Hopf algebra ${\cal A}_R$.
 $ f^j_i $ and $\chi_i$ in eq. (\ref{Tr})   are matrices with c-number
 coefficients.\footnote{This differs from the notation in \cite{ac}
where $ f^j_i $ and $\chi_i$ represent functions on $G_q$.}
 $ f^j_i $ goes to $\delta^j_i$ times the identity matrix, while
$\chi_i$ goes to $\lambda_i$,    in the limit $q\rightarrow 1$, so
that  eq. (\ref{Tr}) reduces to eq. (\ref{Trud}).
 The matrices $ f^j_i $ and $\chi_i$ appear
  in the construction of a bicovariant calculus.
    They are required to satisfy\cite{ac}
\beqa
\Lambda^{nm}_{ij} f^i_k f^j_\ell = f^n_i  f^m_j \Lambda^{ij}_{k\ell}
 \label{frdla}   \\
\chi_i\chi_j-\Lambda^{km}_{ij}\; \chi_k    \chi_m = c_{ij}^k \;
\chi_k \label{sigalg} \\          \Lambda^{ij}_{k\ell}  f^n_i\chi_j  =
 \chi_k f^n_\ell     \\
c_{mn}^i f^m_j f^n_k +f^i_j \chi_k = \Lambda^{mn} _{jk}\chi_m f^i_n
  +c_{jk}^\ell f^i_\ell
\label{id8}
 \eeqa
The second equation states that  $\{ [\chi_i]^a_{\;\;c}\}$
\footnote{When writing the elements of a matrix, it is understood that
the left most
index labels the rows, while the right most one labels the columns.}
defines a
 matrix representation  (the fundamental representation) of
   a  deformed
 Lie algebra.   (From (\ref{rr}) this is not, in general,
  the   algebra generated by the $r_i$'s.  Rather, as we
shall show shortly,    it is the
   algebra generated by the analogue of the left generators $\ell_i$'s.)
Eqs. (\ref{frdla}-\ref{id8}) can be viewed as defining
 an operator algebra for
 $\chi_i$  and $f^i_j$.   In addition to the assumption of the
existence of a fundamental matrix representation for
 $\chi_i$  and $f^i_j$, eqs. (\ref{yb}-\ref{id4}) show that
that there is another  representation (the adjoint representation)
  for    this algebra.  It is obtained by setting
$\chi_i={}^{adj}\chi_i$ and $f^i_j={}^{adj}f^i_j$, where
\be [^{adj}\chi_i]^{\;\;j}_k = c^j_{ki} \qquad
[^{adj}f^i_j]^{\;\;k}_\ell
 = \Lambda^{ik}_{\ell j}  \label{adj} \ee

   With the help of the bicovariant calculus,
i.e. the conditions (\ref{yb}-\ref{id4}), (\ref{frdla}-\ref{id8}) and
the
Yang-Baxter equation for $R$, one can show that the algebra generated by
$T^a_b$ and $ r_i$ is consistent with an associative algebra.       For
example, if we consider the trinomial $Tr_ir_j$ (for simplicity, we
suppress the matrix indices on $T$), by repeated usage of (\ref{rr}) and
(\ref{Tr}), $T$ can be commuted all the way to the right in two
different
ways:
\beqa
Tr_ir_j   &=& (r_kTf^k_i  +T\chi_i )r_j    \cr
               &=   & r_k(r_\ell T f^\ell_j +T\chi_j) f^k_i
         +(r_\ell T f^\ell_j +T\chi_j) \chi_i \label{trr1} \\
         &=   &(\Lambda^{m n}_{\ell k} r_nr_m + c^m_{\ell k} r_m) T
           f^\ell_j  f^k_i  + r_kT \chi_j f^k_i
 + (r_k T f^k_j +T\chi_j)\chi_i
                       \nonumber       \eeqa
\beqa
Tr_ir_j   &=& T(\Lambda^{\ell m}_{ji}r_mr_\ell + c_{ji}^mr_m) \cr
                &=& \Lambda^{\ell m}_{ji} (r_kT f^k_m +T\chi_m)r_\ell +
          c^m_{ji} (r_kT f^k_m +T\chi_m)
          \label{trr2} \\
&=& \Lambda^{\ell m}_{ji} r_k (r_nT f^n_\ell +T\chi_\ell)f^k_m
   +     \Lambda^{\ell m}_{ji} (r_nT f^n_\ell +T\chi_\ell)\chi_m
    +   c^{ m}_{ji} (r_nT f^n_\ell +T\chi_\ell)      \nonumber \eeqa
Equating the right hand sides of
 (\ref{trr1})  and (\ref{trr2}) leads to no new conditions.
Rather,   (\ref{trr1})  and (\ref{trr2}) are identical upon
using the bicovariance conditions (\ref{frdla}-\ref{id8}).

Another ingredient for a bicovariant calculus is
 the analogue of the adjoint matrix elements $M^{\;i}_j$ which now have
  values in $Fun_q(G)$.
 Its coproduct is $\Delta(M_i^{\;j}) = M_i^{\;k} \otimes M_k^{\;j}$.
  The analogue of relation  (\ref{defM}), i.e.
\be \chi_j T = T\chi_i \; M^{\;i}_j \label{defMd} \ee
   can  be
 adapted,  only now the ordering is important,
 and furthermore one has\cite{ac}
\be  M^{\;j}_i f^i_kT = T f^j_i  M^{\;i}_k  \label{MfT}         \ee
These relations can be utilized in obtaining
 the commutation properties for $M^{\;j}_i $:
\beqa  \Lambda^{ij}_{k\ell} M_i^{\;m}M_j^{\;n}   - M_k^{\;i}M_\ell^{\;j}
\Lambda^{mn}_{ij}     &=& 0              \\   M^{\;j}_i r_k
 - \Lambda^{mj}_{nk}  r_m  M^{\;n}_i  &=& c^j_{\ell k}  M^{\;\ell}_i
\label{Mell}
\eeqa     generalizing (\ref{MM}) and (\ref{Mr}).
They are the commutation relations  (\ref{RTT}) and (\ref{Tr})
written in the adjoint representation  (\ref{adj}).
Now  define the analogue of the left generators on the group
  $\ell_i=r_jM^{\;j}_i $.   From  (\ref{defMd}) and (\ref{MfT}),
we get \be T\ell_i - \ell_j f^j_i T = \chi_i T \label{Tell} \ee
generalizing (\ref{Tellud}).
   It is easy to check that $\ell_i$
 commute with $r_j$ as in the undeformed case:
\beqa \ell_i r_k  &=&r_jM^{\;j}_i  r_k =
 r_j\;(\Lambda^{mj}_{nk}  r_m  M^{\;n}_i + c^j_{n k}  M^{\;n}_i)    \cr
 &=& (\Lambda^{mj}_{nk} r_j  r_m  + c^j_{n k} r_j )  M^{\;n}_i
  =r_k r_n M^{\;n}_i    = r_k  \ell_i  \eeqa
where we used (\ref{rr}) and (\ref{Mell}).  The generalization of
 commutation relations (\ref{llud}) can then be obtained
\beqa \ell_i \ell_k &=& \ell_i r_j M^{\;j}_k   = r_j  \ell_iM^{\;j}_k
   = r_j  r_\ell M^{\;\ell}_i   M^{\;j}_k    \cr
   & =&  (\Lambda_{\ell j}^{mn} r_n  r_m  + c_{\ell j}^{n} r_n )
    M^{\;\ell}_i   M^{\;j}_k      \cr
  & =&   \Lambda^{\ell j}_{ik} r_n  r_m  M^{\;m}_\ell   M^{\;n}_j
   + c_{\ell j}^{n} r_n  M^{\;\ell}_i   M^{\;j}_k      \cr
 &=&  \Lambda^{\ell j}_{ik}\; \ell_\ell  \ell_j + c_{ik}^j \;\ell_j
 \label{ll}                    \eeqa
 where we again used (\ref{cMM}) which remains valid in the deformed
case.  We also note that
(\ref{cMM}) corresponds to (\ref{defMd}) written in
the adjoint representation.  From (\ref{ll}), the left generators
satisfy
the same commutation relations  as the $\chi_i$'s in (\ref{sigalg}).
The $\chi_i$'s therefore provide a matrix representation for the
$\ell_i$'s.  It is the fundamental representation and thus irreducible.

The $\ell_i$'s (alone)
 generate a non cocommutative Hopf algebra ${\cal A}_L$ which
 is,    in general, distinct from the Hopf
   algebra ${\cal A}_R$ generated by    the $r_i$'s.  However,
  their centers, if they exist,
coincide.  For example, if one of the generators of ${\cal A}_L$,
say $\ell_0$, belongs to the center of ${\cal A}_L$, then
it also belongs to the center of ${\cal A}_R$.  To see this,
we note  by Schur's lemma that the fundamental matrix representation
 $\chi_0$ of
 $\ell_0$ is a multiple of the identity.  Then  by (\ref{defMd})
\be T \chi_0  =  \chi_0 T = T \chi_j  {M^{\;j}_0}  \ee
Assuming linear independence of the $\chi_i$'s,
  $ {M^{\;j}_0}$  is just $\delta ^j_0$ times the  unit $I$ of the
algebra $G_q$
  (which in fact we assume to be the same as the unit of the full
algebra ${\cal B}$).
This then implies that $ \ell_0 =r_jM^{\;j}_0 =  r_0 $, and since all
$\ell_i$'s commute with elements of   ${\cal A}_R$,  $r_0 $ must be in
its center.   Similarly, we can show that if $ C_L$ is
 a quadratic Casimir for  ${\cal A}_L$, it is also a
  quadratic Casimir for  ${\cal A}_R$.  In this regard, let us write
 \be C_L  = h^{ij} \ell_i\ell_j \label{LCas} \ee
  $h^{ij}$ being c-numbers, and unlike in the undeformed case, $h^{ij}$
    is not,  in general, symmetric.    Then  by Schur's lemma
  $h^{ij} \chi_i\chi_j  $  is a multiple of the identity, and
\be T \;  h^{ij} \chi_i\chi_j  = h^{ij} \chi_i\chi_j \; T =
 h^{ij} T \chi_\ell \chi_k  M^{\;\ell}_i  M^{\;k}_j   \ee
This is solved by
 \be h^{\ell k} I =h^{ij}   M^{\;\ell}_i  M^{\;k}_j   \ee
We can use this to write $C_L $  as a binomial in $r_i$ as follows:
\beqa C_L& =&  h^{ij} \ell_i r_k  M^{\;k}_j
           =   h^{ij} r_k \ell_i M^{\;k}_j    \cr
         & =&  h^{ij} r_k r_\ell  M^{\;\ell}_i M^{\;k}_j
           =   h^{\ell k} r_k r_\ell   \eeqa  Since (\ref{LCas})
          commutes with $r_i$,
$C_R  =  h^{\ell k} r_k r_\ell $ is  a Casimir for ${\cal A}_R$.
It is a natural choice for the Hamiltonian  of a q-rigid rotor.
The above procedure can be repeated for any higher order Casimirs.

Finally, let us
enlarge our algebra ${\cal B}$ to include
a   bimodule $\Gamma$.  One then assumes the existence
of an exterior derivative $d$ which maps  $G_q$ to $\Gamma$, and
following\cite{bico} satisfies  the usual Leibniz rule.
We next introduce the left and right invariant one forms.  We define
them as in (\ref{lrif}).  They
obey the Maurer-Cartan equations:
\beqa  d   \omega_L  +  \omega_L \wedge \omega _L   & =& 0     \cr
       d   \omega_R  -  \omega_R \wedge \omega _R   & =& 0  \label{mc}
         \eeqa
If we postulate that $dT$ has the same commutation relations with
 $\ell_i$ and $r_i$
as does $T$, it then  follows that $\omega_L$ commutes with
$\ell_i$, and $\omega_R$ commutes with $r_i$, as in the undeformed
case.    For the latter,  multiply
$     d T    r_i - r_j dTf^j_i=   d T  \chi_i $ on the right by
 $T^{-1}$
and  $     r_i  T^{-1} - T^{-1}   r_j   T  f^j_i   T^{-1}=    \chi_i
 T^{-1} $  on
the left by $dT$, and then take the difference to find
 \be [\omega_R \; ,\; r_j] \;Tf^j_i T^{-1}  = 0  \ee
   The result that  $\omega_R$ and $ r_j$ commute  follows
using the existence of an antipode   $\kappa'(f^k_j) \cite{ac}$, where
$\kappa'(f^k_j)\; f^j_i = f^k_j\; \kappa'( f^j_i) = \epsilon'\;
\delta^k_i $.
$\epsilon'$ is the co-unit, which for us is just the unit matrix in
the defining representation.
The nonvanishing commutation relations of the generators
 $\ell_i$ and $r_i$       with
the left and right invariant one forms are
\beqa
 [\omega_L\; ,\; r_j \; \kappa'(f^j_i) ] &=& [\omega_L\;,\;
 \chi_j \; \kappa'(f^j_i) ]       \cr
 [\omega_R\; ,\; \ell_j\; f^j_i ] &=& - [\omega_R\;,\; \chi_i]    \eeqa
Upon writing $\omega_L = \omega_L^i \;\chi_i$  and
             $\omega_R = \omega_R^i \;\chi_i$,  they can be  rewritten
\beqa           [\omega_L^i,  r_j]_q \equiv
\;\omega_L^i  r_j  -\Lambda^{ki}_{\ell j}\; r_k \omega_L^{\ell}
&=& c^i_{kj}\; \omega_L^k
      \cr      [\ell_j,\omega_R^i]_q \equiv
\;\ell_j \omega_R^i  -\Lambda^{ki}_{\ell j}\; \omega_R^{\ell}  \ell_k
&=& c^i_{kj}\; \omega_R^k       \label{crowlr}   \eeqa
generalizing (\ref{crowlrud}).
Once again, the components $ \omega_L^i$ and $\omega_R^i$ are related by
(\ref{rbos}).  To derive the second equation in
(\ref{crowlr}) one can  substitute (\ref{rbos}) into
$[    \omega_L^i , \ell_j] =0$.

This system suggests a natural generalization to  gauge
theories based on quantum groups.  This was  pursued
in \cite{Cas},\cite{us}  from the
Lagrangian perspective.
\footnote{Gauge theories based on quantum groups were also
studied on the lattice. \cite{Fr},\cite{bsv}
The continuum limit for these theories is nontrivial.  On the other hand,
in \cite{Cas},\cite{us} and what follows one works directly on the
 continuum.}
Here we view the theory from the Hamiltonian perspective.
For this one can now drop the zero curvature conditions  (\ref{mc}),
and let $\omega_L$ (or $\omega_R$) denote arbitrary connection one forms.
From (\ref{crowlr}),   $r_i$ (or $\ell_i$)  generate global
transformations.  Following Castellani\cite{Cas}, if we introduce
 `infinitesimal' parameters $\epsilon^i$, which are not c-numbers,
infinitesimal variations of the one forms $\omega^i_L$ can be expressed by
\be \delta \omega_L^i = - [\omega_L^i ,  r_j ]_q \;\epsilon^j =
- c^i_{kj}\; \omega_L^k  \epsilon^j     \ee
A natural question concerns the generators of local or gauge
transformations, i.e.  the analogue  of the Gauss law constraints.
From \cite{Cas}, local quantum group transformations are of the form
\be \delta \omega_L^i =
 -d \epsilon^i - c^i_{kj}\; \omega_L^k  \epsilon^j  \;, \label{qgtran}\ee
where  the  `infinitesimal' parameters $\epsilon^i$ are now functions
on space-time, which is considered here as a commuting manifold.   The
Gauss law generators ${\cal G}_i$ and space components $A_\mu^i$ of
 $\omega_L^i$   are also functions on space-time.
  The equal time commutator of Gauss law generators with
           $ A_\mu^i$    should now include a central term
\beqa         [A_\mu^i (x), {\cal G}_j(y)]_q  &\equiv  &
\;A_\mu^i (x) {\cal G}_j (y)  -\Lambda^{ki}_{\ell j}\; {\cal G}_k (y)
 A_\mu^{\ell}(x)          \cr
&=& c^i_{kj}\; A_\mu^k(x)\delta(x-y) + \delta^i_j
\frac{\partial }  {\partial x^\mu }   \delta(x-y)     \label{crag}
    \eeqa        The Gauss law constraints  should form a closed
algebra if they are to be first class.   A natural choice is
  a q-Lie algebra at every point in space:
\be {\cal G}_i(x) {\cal G}_j(y) -\Lambda^{mn}_{ji} {\cal G}_n(y)
{\cal G}_m(x)= c^k_{ji} {\cal G}_k(x) \delta(x-y)\;.\label{gcr}  \ee
However, the
 algebra generated by $A_\mu^i$ and ${\cal G}_j$
  is, in general, nonassociative.     In fact, the algebra
for the Gauss law generators alone is, in general, nonassociative.
On the other hand, it can be checked that associativity is recovered
 for the so-called `minimal'
deformations\cite{Cas}\cite{us},
 where   \be \Lambda^{ij}_{k\ell}  \Lambda^{k\ell}_{mn}
=\delta^i_m \delta^j_n \label{mnmlty} \ee
One can  wonder if the Gauss law can be expressed in terms of
some analogue of the electric fields $\pi_i^\mu$, similar
to what is done in the underformed case.
  This should at least be possible in the
minimal case.  We thus write:
\be {\cal G}_i(x) = - \frac{\partial }  {\partial x^\mu }
\pi_i^\mu (x)  +    c^k_{ji} \pi_k^\mu (x) A^j_\mu(x)  \ee
To recover the derivative
  term in (\ref{crag}) one then needs
\be    [A^i_\mu(x),\pi_j^\nu(y)]_q \equiv
A^i_\mu(x) \pi_j^\nu(y) - \Lambda^{ki}_{\ell j} \pi^\nu_k(y)
  A^\ell_\mu(x) = \delta^i_j\; \delta^\nu_\mu \;\delta(x-y) \ee
giving a deformation of the canonical commutation relations.
For the remaining term in (\ref{crag}) one needs
\be c^m_{\ell j} \Lambda^{ni}_{km}\; A^k_\mu (x) A^\ell_\nu (y) =
    c^n_{m k} \Lambda^{ki}_{\ell j}\; A^m_\nu (y) A^\ell_\mu (x)   \ee
This is indeed valid for the minimal case if one  chooses the
 following  commutation relations for the gauge potentials:
\be  A^k_\mu (x) A^\ell_\nu (y) =
    \Lambda^{k\ell}_{m j}\; A^m_\nu (y) A^j_\mu (x)   \ee
which agree  with those found in \cite{Cas}.   Concerning the
commutation relations between the electric fields, this can be
obtained (for minimal deformations) by again appealing to associativity.
The algebra of the operators  $A^i_\mu$ and $\pi_j^\nu$ is associative
for
\be \pi_i^\mu (x)    \pi_j^\nu (y) = \Lambda_{ji}^{k\ell}
                          \;      \pi_\ell^\nu (y)   \pi_k^\mu (x) \ee

From the above discussion, it may
appear that gauge theories based on quantum groups
are  possible only for `minimal' deformations.
This is  consistent with the result found in \cite{Cas} that the set of
gauge transformations (\ref{qgtran}) closes only for such deformations.
For us, minimality was
needed to make the algebra generated by ${\cal G}_i(x)$, satisfying
 (\ref{gcr}), be associative.
On the other hand, (\ref{gcr}) may be too simple a generalization of
the global symmetry algebra defined by (\ref{rr}).
  A less restrictive choice would have space-dependent braiding
matrices, R-matrices and structure constants.  More precisely,
${\cal G}_i(x) $   can be regarded as `right' generators having
both a continuous and discrete index, whose corresponding
braiding matrices $\tilde\Lambda$  and structure constants $\tilde c$
 have the simple form
\be \tilde\Lambda^{(mz)\;(nw)}_{(jy)\;(ix)}    =\Lambda^{mn}_{ji}\;
\delta(x-z)\;\delta(y-w)  \qquad \tilde c^{(kz)}_{(jy)\;(ix)} = c^{k}_{ji}
\;\delta(x-z)\;\delta(x-y)  \ee This choice is consistent with the
 bicovariance  conditions  if (\ref{mnmlty}) is satisfied.
Whether or not one can have a more general expression for $\tilde\Lambda$
 and  $\tilde c$ would be of interest.   Moreover, it is perhaps
also too restrictive to require that space-time defines
 a normal manifold, when this is not the case for the target manifold.
 Possibly, a more general class of gauge theories based on quantum groups
  can be developed if one is able to drop this requirement as well.

   \section*{Acknowledgments}

A.S. thanks S. Frolov for discussions and
 the members of Dipartimento di Scienze Fisiche,
Universit\`a
di Napoli, for their  hospitality and support where this work was
initiated. A.S. was  supported in part by the U.S. Department of Energy
under contract number DE-FG02-96ER40967. This work was partially
supported
by PRIN 97 ``SIN.TE.SI".

\end{document}